\documentstyle[12pt,aasms4]{article}
\begin{document}

\title{The multi-phase nature of three intracluster media }

\author{Massimiliano~Bonamente$\,^{1}$, and Richard~Lieu$\,^{1}$}

\affil{\(^{\scriptstyle 1} \){Department of Physics, University of Alabama,
Huntsville, AL 35899, U.S.A.}\\
}

\begin{abstract}
Among the models proposed to account for the new
component of diffuse EUV and soft X-ray emission from clusters of galaxies
(first discovered in Virgo [1])
are two key contestants: the non-thermal scenario which postulates
a population of relativistic electrons undergoing inverse-Compton (IC)
interaction with the cosmic microwave background [2,3], and the original
conjecture [1] that the radiation is from a thermal warm gas at 
a temperature of $\sim$ 10$^{5-6}$ K.  Currently a consensus set of
limiting values on
cosmological parameters
favor the
thermal gas interpretation [4].
We also argued, based on pressure balance
within the intracluster medium (ICM), 
that the non-thermal approach has formidable
difficulties [5].
Here we describe a
spatial analysis of the soft X-ray excess emission of three clusters (Virgo,
A2199, and Coma), using archival ROSAT/PSPC data, which reveals resolved
features of
cold intracluster clouds in absorption spreading over vast distances.
Within the sample there is
good indication that the soft excess radial trend (SERT, which qualitatively
means a rising importance of the soft component with cluster radius) is
due to a centrally peaked distribution of cold matter, with Coma having
the least effect and no direct evidence for absorption.  The data strongly
suggest an intermixed ICM which contains gas masses at a wide
range of temperatures, and the
soft excess is due to a warm intermediate phase.
\end{abstract}

In an accompanying work
we found several pieces of evidence, based mainly on the
detection of cloud silhouettes in the EUV, that the ICM of the cluster A2199
is multi-phase [5].  This {\it Letter} presents X-ray (0.2 - 2.0 keV)
data, taken by the PSPC, of a cluster sample which provide
independent results pointing to a generally intermixed ICM with important
roles played by gas phases at temperatures lower than that of the
hot (virial) gas.

We begin with Virgo, and show in Figure 1 the SERT in the 1/4 keV band
with three noteworthy points.  Firstly,
the line-of-sight HI column density (N$_H$)
was shown by a recent 21 cm measurement [1] to radially increase from
N$_H =$ 1.8 $\times$ 10$^{20}$ cm$^{-2}$ at the cluster center to
N$_H =$ 2.0 $\times$ 10$^{20}$ cm$^{-2}$ at the radius interval of 
15 -- 19 arcmin.
This was confirmed by IRAS 100 $\mu$m images [6], 
consequently in Figure 1 we
already took its effect on the SERT into account.  Secondly, given
the radial HI gradient (which continues its rising trend beyond
19 arcmin), and the known anti-correlation between HI and the 1/4 keV
diffuse sky background [7], one must assess how much the PSPC background
was underestimated when it was determined, as in our case, from a
$\sim$ 40 -- 50 arcmin annulus
centered at M87.  
Of most concern are the 10 --15  and 15 -- 19 arcmin annuli,
where the 1/4 keV sky background accounts for 12 -- 21\% of the detected 
flux in this band.  A re-scaling of this background in accordance with
the HI gradient [7]  over the
corresponding regions
only leads to a negligible effect
on the 1/4 keV excess (viz. a reduction by 1 -- 2\% from our reported values
of 30 -- 40 \% excess).  Thirdly, a statistically significant rising
SERT was also revealed by our recent EUVE (0.069 - 0.19 keV)
observation of Virgo, which featured an {\it in situ} background
measurement by means of the offset pointing technique detailed earlier [8].
It is however the higher signal-to-noise PSPC data which enable us to probe
the nature of the soft emission using
image diagnostics.

We provided facts which form a compelling case for 
interpreting the SERT as due, at least in part, to intracluster
absorption by an even cooler phase [5].   This {\it Letter} explains why we
are confronted with the reality of widespread absorption -
the PSPC has already resolved the effect into small clouds distributed
throughout the ICM.  Such an inference was made after
evaluating the smoothness of the 1/4 keV excess image (for details
on the procedures used to obtain this image, see the caption of
Figure 1).  Specifically
the presence or not of deviations in the spatial distribution
of signals from Poisson behavior was assessed.  As a control
experiment, we initially applied the test to a `blank field', acquired during
a PSPC pointed observation of the (undetected) UV star beta Leonis,
when the field was not illuminated by any
source other than the sky background.
When this background was
subtracted in the normal manner (i.e. using an outer field annulus
to determine it, and correcting for vignetting effects before
applying to another part of the detector)
the resulting spread of significances follows, as expected, a 
gaussian of null mean
value and $\sigma$= 1, see Figure 2.

The same method
was then applied to three annuli of Virgo.  The
results, displayed in Figure 3, indicate the presence of spatial
structures in Virgo's soft emission.  The soft excess is clearly
revealed by the positive mean value at all radii.  However, even
if the data are fitted with a $\sigma$=1 gaussian of variable mean,
the agreement remains unsatisfactory, due to residuals at negative
$\sigma$, which can only be interpreted as signatures of absorption
at scale lengths $\leq$ the Point Spread Function (PSF) of the PSPC,
where the soft component is silhouetted by cold clouds along the
line-of-sight.  Starting from the cluster center (0 -- 4 arcmin
from M87), we found on the $- \sigma$ side
a deviation from the expected (best-fit)
gaussian by $+ \sim$ 11 \% 
in the total number of scanned regions.
The effect decreases with radius, since the same percentage deviation
reduces to $+ \sim$ 4 \% in the 4 -- 7 arcmin annulus, and further
out there is
no longer any evidence for non-gaussian behavior.  This radially declining
influence of absorption - an explanation  
of the SERT - is naturally understood in terms of a centrally condensed
distribution of cold gas.
In Figure 4 
we show the three deepest absorption features which exist in the central
area, positionally coincident with prominent radio lobes [9] and with
locations where the hot ICM has a lower reported temperature [10].

We proceed to the next cluster of our sample, A2199, where again
we focus on the PSPC image.  At EUV energies a strongly rising
SERT was found for this cluster [5].  Yet the same is not true
in soft X-rays, as is shown in Figure 5 where it can be seen that
while the center exhibits flux depletion, analogous to
the EUV, the outer radii are not associated with a soft excess,
nor with a rising trend.  Could absorption have played a role in this
large scale radial behavior ?  Our simulations indicate that that it
is indeed possible to compare and contrast the EUV with the soft X-rays
in terms of a warm component which has
a large intrinsic EUV to soft X-ray flux ratio, coupled with
Galactic {\it and} an appropriate amount of intracluster absorption
(the latter with N$_H$ between a few $\times$ 10$^{19}$ and 10$^{20}$
cm$^{-2}$)
along the line-of-sight.  The outcome is an absorbed flux which remains
within the sensitivity of the EUVE observation, but evades detection
by the PSPC.  

Such a scenario is put to test by spatially analyzing the
distribution of the soft excess of A2199, Figure 6.  The central region
corresponds to a gaussian of expected width but negative mean,
symptomatic of a large absorption area wherein the clouds are
unresolved [5].  As one moves towards the outer radii the best
gaussian mean shifts towards {\it positive} values while the $\sigma =$ 1
width does not fit the left half, where a `tail' of resolved
absorption clouds is evident.  
This `tail' biases the
data mean towards negative values, producing the illusion of an overall
depletion in soft X-rays when there actually is an {\it excess} flux.
The discovery of absorbing clouds then applies at least
out to a radius of 10 arcmin
($\sim$ 0.4 Mpc for H$_o =$ 75), implying an area $\sim$ 25 times
larger than that of the cooling flow, and $\sim$ 3 times larger than
that of the central EUV `shadow' [5].

Equally revealing are the PSPC data of Coma, our last cluster,
as it provides more independent scrutiny of the intermixed model.
This cluster has a weak (i.e. nearly flat) SERT, although
there is a soft X-ray excess at all radii [11], see Figure 7.
According to
our proposed interpretation, then, the ICM of Coma would probably not
be as subject to absorption effects as the other clusters.  This
is confirmed by the spatial distribution of the soft excess,
which shows no evidence for deviation from a smooth
(i.e. gaussian) behavior in any annulus, Figure 8.
The absence of a cooling
flow in Coma may be the reason why there is less cold gas, although
such an explanation does not account for the detection of absorption
in A2199 at radii of 7 -- 10 arcmin.  A more plausible idea is that
the generation of a cold phase does not proceed at high rates
when the hot ICM has an unusually high temperature, as is the case for Coma.

In conclusion, analysis of PSPC images of three clusters revealed that
the two which exhibit a strong SERT (Virgo and A2199) also have
a widespread distribution of absorbing clouds, rendering the prospect
of interpreting the SERT as an absorption effect attractive.
Typical values for the mass and column density of the cold gas,
as estimated from the data, are respectively 
$\sim$ 5 $\times$ 10$^{10}$ M$_{\odot}$ Mpc$^{-3}$ and
$\sim$ a few $\times$
10$^{19}$ cm$^{-2}$ [5].   The soft excess takes the form of a hitherto
unresolved glow of diffuse emission filling the ICM, since 
(unlike absorption) there is
no evidence in the PSPC images for isolated emission `blobs'.  Thus the
possibility of a warm intermediate phase which has larger filling
factor than the cold phase has also become attractive.  Certainly
one can no longer continue with the notion of the soft excess as
a systematic effect of some kind
[12,13]:
if the gaussian means can be centered
at zero (rather than their currently
positive values) as a result of correcting such effects, one must
face the absurdity of interpreting the `tails' at negative $\sigma$ as
absorption of null signals.

\newpage

\noindent
{\bf References}

\noindent
~1. Lieu, R., Mittaz, J.P.D., Bowyer, S., Lockman, F.J.,
Hwang, C. -Y., Schmitt, \\
\indent  J.H.M.M. 1996a, \it Astrophys. J.\rm, {\bf 458}, L5--7. \\
\noindent
~2. Hwang, C.-Y. 1997, {\it Science}, {\bf 278}, 1917. \\
\noindent
~3. Sarazin, C.L., Lieu, R. 1998, {\it Astrophys. J.}, {\bf 494}, L177--180. \\
\noindent
~4. Cen, R. and Ostriker, J.P. 1999, {\it ApJ}, {\bf 514}, 1-6. \\
\noindent
~5. Lieu, R., Bonamente, M. and Mittaz, J.P.D 2000, {\it Nature} submitted. \\
\noindent
~6. Wheelock et al. 1994, {\it IRAS Sky Survey Explanatory Supplement}, (JPL Publication \\
\indent 94-11), Pasadena, CA. \\
\noindent
~7. Snowden, S.L., Egger, R., Finkbeiner, D.P., 
Freyberg,M.J. and Plucinsky, P.P. 1998, \\
\indent {\it Astrophys. J.},{\bf 493}, 715. \\
\noindent
~8. Lieu, R., Bonamente,M. ,Mittaz, J.P.D., Durret, F., Dos Santos, S. and \\
\indent Kaastra, J.S.  1999, {\it ApJ}, {\bf 527}, L77. \\
\noindent
~9. Harris, D.E., Owen, F., Biretta, J.A., 
and Junor, W. 1999, {\it Proceedings of \\
\indent the Workshop `Diffuse thermal 
and relativistic plasma in galaxy cluster'}, Ringberg \\
\indent Castle Germany, MPE report 271, 111. \\
\noindent
~10. Boehringer, H. 1999 , {\it Proceedings of
the Workshop `Diffuse thermal and relativistic \\ 
plasma in galaxy cluster}, Ringberg
Castle Germany, MPE report 271, 115. \\
\noindent
~11. Lieu, R., Mittaz, J.P.D., Bowyer, S., Breen, J.O., Lockman, F.J.,\\
\indent Murphy, E.M. and Hwang, C.-Y. 1996, {\it Science}, {\bf 274}, 1335. \\
\noindent
~12. Arabadjis, J.S. and Bregman, J.N. 1999, {\it Astrophys. J.}, {\bf 514}, 607. \\
\noindent
~13. Bowyer, S., Berghoefer, T.W and Korpela, E.J. 199, {\bf Astrophys. J.},\\
\indent {\bf 526}, 592. \\
\noindent
~14. Mewe, R., Gronenschild, E.H.B.M. and van den Oord, G.H.J. 1985, {\it A \& A}, {\bf 62}, 197 .\\
\noindent
~15. Mewe, R., Lemen, J.R., and van den Oord, G.H.J. 1986, {\it A \& A}, {\bf 65}, 511 .\\
\noindent
~16. Kaastra, J.S. 1992 in \it An X-Ray Spectral Code for OpticallyThin Plasmas \rm \\\indent
(Internal SRON-Leiden Report, updated version 2.0). \\
\noindent
~17. Kaastra, J.S., Lieu, R., Mittaz, J.P.D., Bleeker, J.A.M., Mewe, R., Colafrancesco, S. \\
\indent and Lockman, F.J. 1999, {\it Astrophys. J.}, {\bf 519}, L119.\\
   
\vspace{3mm}

\noindent
{\bf Figure captions}

\noindent
Figure 1: The SERT effect of the Virgo cluster, illustrated by a plot
against cluster radius of the soft X-ray fractional excess $\eta$,
defined as $\eta = (p - q)/q$, where for a given
annulus $p$ is the observed 1/4 keV
band (defined here as PSPC PI channels 18-41, or $\sim$ 0.2 -- 0.4 keV) 
flux
after subtracting the sky background, and $q$ is the
expected flux from the hot ICM as determined by fitting the
PI channels 50 -- 200 ($\sim$ 0.5 -- 2.0 keV)
using the MEKAL thin plasma emission code 
[14 - 16]
and Galactic absorption as described in the text.
Note that the same subtraction of
background and hot ICM contribution
was applied, except to individual regions
rather than entire annuli, when we investigated the spatial distribution
of the 1/4 keV band excess in Figures 2, 3, and 5.

\noindent
Figure 2: A statistical test of the small scale smoothness of a typical
PSPC 1/4 keV sky background in a `blank field' observation, using a region
of the detector $\sim$ 20 arcmin off-axis.  The `background' was
determined from another region $\sim$ 40 arcmin off-axis, and
was subtracted from the first region after vignetting correction.
The spatial distribution of the resulting signals were sub-divided
equally into small boxes of size 0.5 arcmin $\times$ 0.5 arcmin,
and a histogram is plotted to show the number of occurences above and below
a mean value of zero in units
of $\sigma$, the standard deviation of each box obtained by adding in
quadrature the respective Poisson errors in the measured flux and the
subtracted component.  The box size is larger than the PSPC PSF at all
energies, and encloses sufficient counts to ensure that one is in
the gaussian limit.  The best-fit gaussian (dashed line) is obtained
by varying only the mean, to accomodate the possibility of a finite
(i.e. positive or negative) subtracted signal: its width remains fixed at unit
$\sigma$, while
its normalization is determined by the conservation of total box number.
The best mean is fully consistent with zero, and the absence of
fit residuals implies that the test reveals smoothness of the image.

\noindent
Figure 3: A statistical test of the small scale smoothness of the
1/4 keV excess in Virgo.  The three regions of concern were divided
into small boxes as described in Figure 2, the same applies to
the best-fit gaussian.

\noindent
Figure 4: The image of 1/4 keV excess of Virgo, expressed in units
of $\sigma$ above and below the null mean value expected for the
case of no soft excess, where $\sigma$ is defined in Figure 2 and
the box size used for computations is 1.25 arcmin $\times$ 1.25 arcmin.
The cross marks the position of M87.  Pockets of absorption are
evidently embedded in an unresolved glow of soft excess emission.

\noindent
Figure 5: As in Figure 1, except now for the cluster A2199.  The
Galactic line-of-sight HI column density used is  our measured value
of 8.3 $\times$ 10$^{19}$ cm$^{-2}$ [17].

\noindent
Figure 6: As in Figure 3, except now for the cluster A2199.  See also the
information given in Figure 5.

\noindent
Figure 7: As in Figure 1, except now for the Coma cluster.  The
Galactic line-of-sight HI column density used is  our measured value
of 8.7$\times$ 10$^{19}$ cm$^{-2}$ [11].

\noindent
Figure 8: As in Figure 3,  except now for the Coma cluster.  See also the
information given in Figure 7.

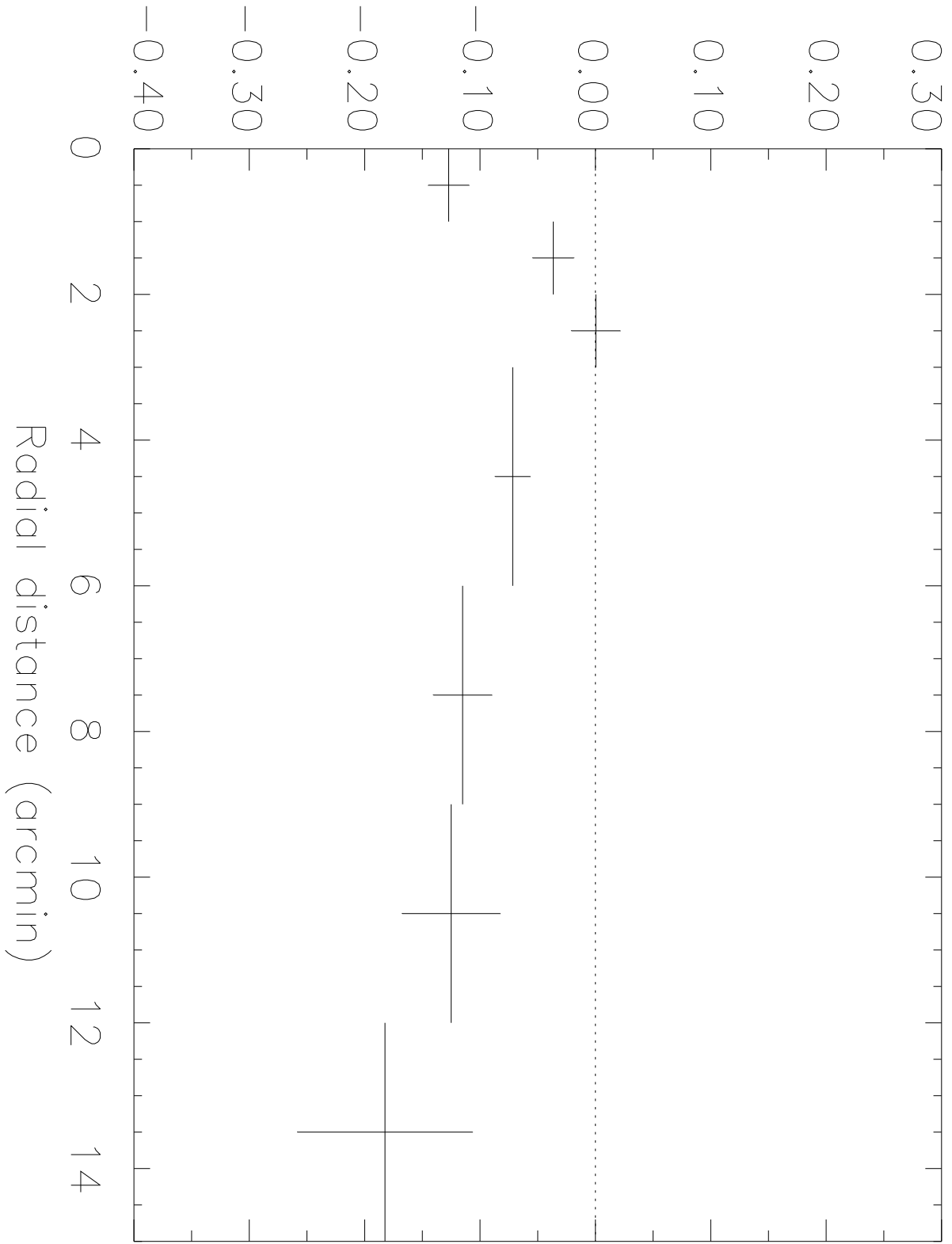
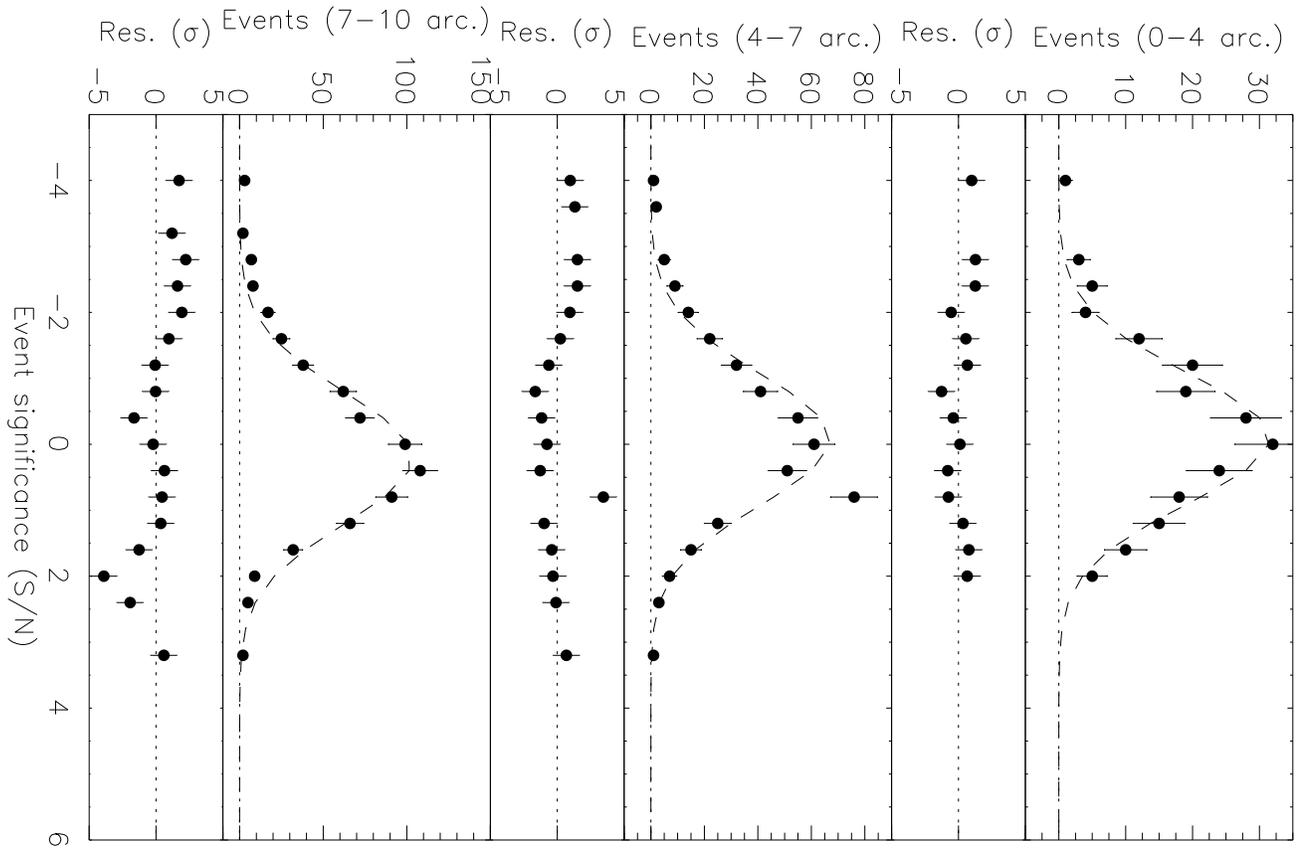
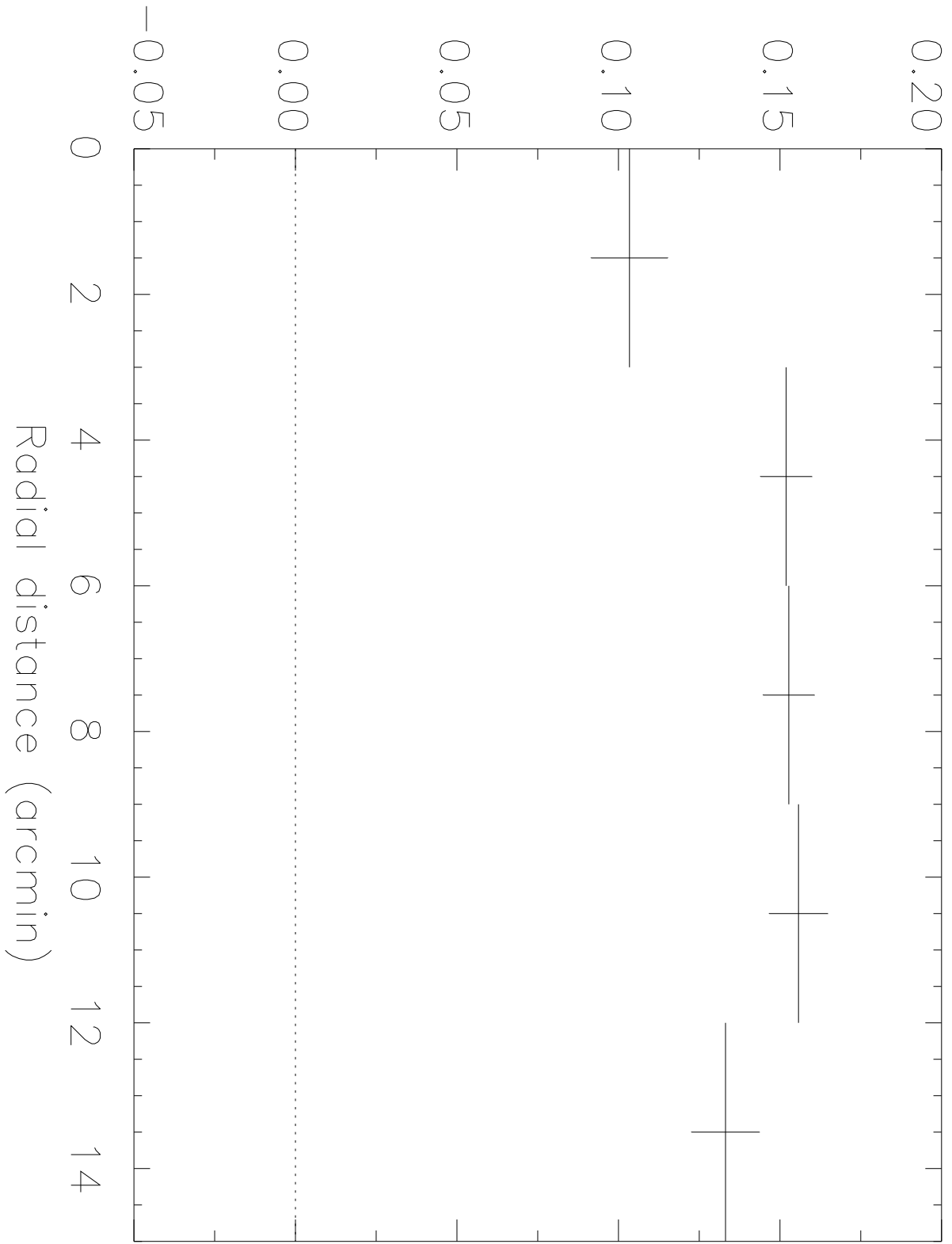
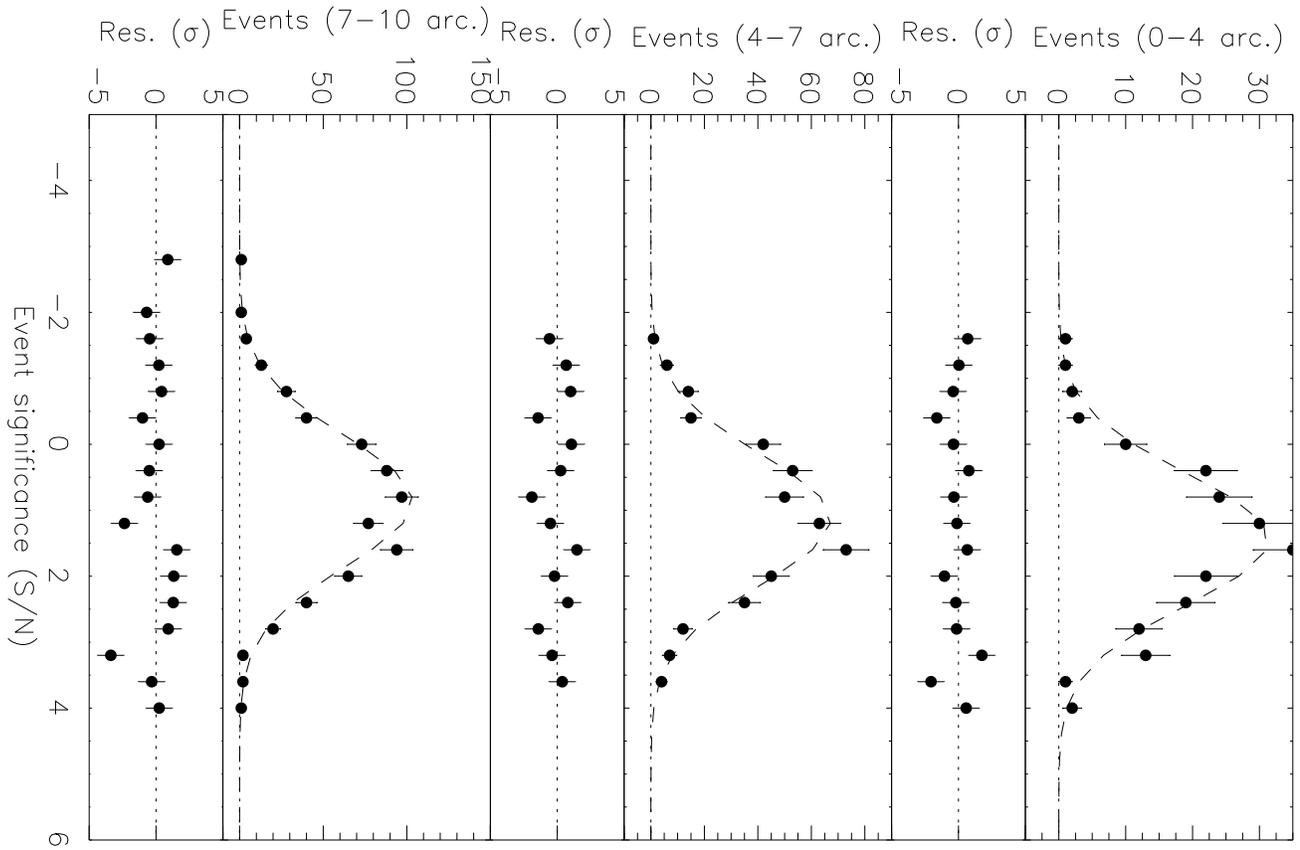
\end{document}